\documentclass[aps,prl,twocolumn,superscriptaddress,showpacs]{revtex4-1}
\usepackage{amsmath}
\usepackage{graphicx}
\usepackage{epsfig}
\usepackage{bm}
\usepackage{hyperref}
\def\be{\begin{equation}}
\def\ee{\end{equation}}
\def\bea{\begin{eqnarray}}
\def\eea{\end{eqnarray}}
\def\bean{\begin{mathletters}\begin{eqnarray}}
\def\eean{\end{eqnarray}\end{mathletters}}

\begin{document}


\title{An Equation of State for Anisotropic Solids under Shock Loading}

\author{Alexander A. Lukyanov}
\affiliation{Abingdon Technology Centre, Schlumberger, Abingdon,
OX14 1UJ, UK}
\begin {abstract}
An anisotropic equation of state is proposed for
accurate extrapolation of high-pressure shock Hugoniot states to
other thermodynamics states for shocked single crystals and
polycrystalline alloys. The proposed equation of state represents
mathematical and physical generalization of the Mie-Gr\"{u}neisen
equation of state for isotropic material and reduces to this
equation in the limit of isotropy. Using an anisotropic nonlinear
continuum framework and generalized decomposition of a stress
tensor [A. A. Lukyanov, Int. J. Plasticity \textbf{24}, 140
(2008)], the shock waves propagation along arbitrary directions in
anisotropic solids of any symmetry can be examined. The
non-associated strength model includes the distortion effect of
the yield surface which can be used to describe the anisotropic
strength differential effect. A numerical calculation showed that
the general pulse shape, Hugoniot Elastic Limits (HELs), and
Hugoniot stress levels for aluminum alloy 7010-T6 agree with the
experimental data. The results are presented and discussed, and
future studies are outlined.
\end{abstract}
\date{\today}
\maketitle
\section{Introduction}
\label{intro}

The shock waves and high-strain-rate phenomena are involved in
many physical phenomena, therefore, we are interested in
understanding the physical properties of solids under these
non-trivial conditions. Investigation of different anisotropic
solids (e.g., single crystals and polycrystalline alloys) under
shock loading has found significant interest in the research
community \cite{Wallace}-\cite{Lukyanov}. Modern, high-resolution
methods to monitoring the stress and particle velocity histories
in shock waves and equipment have been created
\cite{Meyers}-\cite{Stoffel}; numerous investigations into the
mechanical properties of different classes of materials have been
undertaken \cite{Meyers},
\cite{Steinberg}-\cite{BhattachatyaSrivastava}; and numerous
phenomenological as well as microscopic models have been developed
\cite{Andersonetal}, \cite{Meyers}, \cite{Steinberg},
\cite{AsayShahinpoor}, \cite{EliezerGhatakHora}. However, in spite
of a perfectly adequate general understanding, experimental
methodology, and theory, material models do not agree in detail,
especially for anisotropic solids. For many years, it has been
assumed that the response of materials to shock loading is
isotropic, and only recently has anisotropy in the shock response
attracted the attention of researchers. In the early 2000s, Gray
III {\it et al.} \cite{Grayetal1}, \cite{Grayetal2} showed that
under shock loading conditions (one-dimensional strain space), the
variation of the Hugoniot elastic limit (HEL) or the yield
strength of annealed zirconium was consistent with the
quasi-static experimental data. To describe the anisotropic solids
response under shock loading the following several factors must be
addressed during the numerical modelling: (a) large compressions
lead to a nonlinear behavior -- an equation of state (EOS) is
required, (b) low yield stresses lead to large plastic
deformations -- an appropriate strength model is required. To
address these factors, thermodynamically consistent framework for
modelling the response of single crystal and polycrystalline
alloys under shock loading was developed. This framework, building
on the thermodynamic approach of Wallace \cite{Wallace} and
continuum framework of Johnson
\cite{Johnsonetal}-\cite{WineyGupta}, couples nonlinear elasticity
with non-associated anisotropic plasticity within a
thermodynamically consistent numerical incremental formalism
\cite{WineyGupta}.

\section{Shortcomings of the isotropic Equation of State}
\label{sec:1}

The EOS for isotropic materials typically defines the pressure as
a function of density $\rho$ (or specific volume, $\nu$) and
specific internal energy $e$. Experimental shock Hugoniots have
been widely used as reference state data for extrapolating to
other high-pressure and high-temperature thermodynamic states. The
extrapolation has been done by using a very popular form of
equation of state that is used extensively for isotropic solid
continua is the Mie-Gr\"{u}neisen EOS:
\begin{equation}
\label{Luk1_1} \displaystyle\Gamma \left({\nu}\right)=\nu\left(
{\frac{{\partial p}}{{\partial e}}} \right)_{\nu},
\ \ \displaystyle\Gamma (\nu) = \frac{{\gamma_{0} + a\mu} }{{1
+ \mu} },
\end{equation}
where $\Gamma \left( {\nu} \right)$ is the Gr\"{u}neisen gamma,
$\gamma_{0}$ is the initial Gr\"{u}neisen gamma, $a$ is the first
order volume correction to $\gamma_{0}$, and $p$ is the hydrostatic
pressure. Assuming $\Gamma (\nu)$ to be a function of volume only
(\ref{Luk1_1}), the Mie-Gr\"{u}neisen EOS provides a simple
extrapolation of a known set of Hugoniots data to other sets
of thermodynamic states:
\begin{equation}
\label{Luk1_2}
p=f\left({\rho ,e}\right)=P_{H}\cdot\left({1-\frac{{\Gamma
}}{{2}}\mu}\right)+\rho\Gamma e,
\end{equation}
where $P_{H}$ is the Hugoniot pressure, $\displaystyle\mu =
\frac{{\rho} }{{\rho_{0}}}-1$ is the relative change of volume.
The shock compression process is characterized by the
Rankine-Hugoniot jump relations \cite{Meyers}, \cite{Bushmanetal},
\cite{AsayShahinpoor}, \cite{EliezerGhatakHora}. The
Rankine-Hugoniot equations for the shock jump conditions includes
five variables: pressure $p$, particle velocity $u_{p}$, shock
velocity $U_{s}$, density $\rho$ (or specific volume, $\nu$ ) and
energy $e$. Hence, an additional equation is needed. In many
dynamic experiments,   $u_{p}$ (the particle velocity directly
behind the shock) and  $U_{s}$ (the velocity at which the shock
wave propagates through the medium) are measured. The Hugoniot
pressure and a shock velocity $U_{s}$ are a non-linear function of
particle velocity $u_{p}$. Following Steinberg's model
\cite{Steinberg}, the shock velocity is:
\begin{equation}
\label{Luk1_3}\displaystyle U_{s} = c + S_{1} u_{p} + S_{2} \left(
{\frac{{u_{p}} }{{U_{s}}}}\right)u_{p} + S_{3}\left({\frac{{u_{p}}
}{{U_{s}}}}\right)^{2}u_{p} ,
\end{equation}
where $c$ is the velocity curve intercept (can be approximated by
the bulk speed of sound). Therefore, the Mie-Gr\"{u}neisen
equation of state with cubic shock velocity as a function of
particle velocity defines pressure as
\begin{equation}
\label{Luk1_4}\displaystyle p = \left\{{{\begin{array}{*{20}c}
\displaystyle {\frac{{\rho_{0}c^{2}\mu \left[ {1 + \left( {1 -
\frac{{\Gamma}}{{2}}} \right)\mu - \frac{{\Gamma}}{{2}}\mu^{2}}
\right]}}{{\left[ {1 - \left( {S_{1} - 1} \right)\mu - S_{2}
\frac{{\mu^{2}}}{{\mu + 1}} - S_{3} \frac{{\mu^{3}}}{{\left( {\mu
+ 1} \right)^{2}}}} \right]^{2}}}} + \hfill \\
{+\left( {1+ \mu}  \right) \cdot \Gamma \cdot E, \quad  \mu > 0;}
\hfill \\
{\rho_{0}c^{2}\mu + \left( {1 + \mu}  \right) \cdot \Gamma \cdot
E,\quad \mu < 0;} \hfill \\
\end{array}} } \right.,
\end{equation}
\begin{equation}
\label{Luk1_5} \displaystyle c=\sqrt\frac{K}{\rho_{0}},
\end{equation}
where $E$ is the internal energy per initial specific volume
$\displaystyle\left(E=\frac{e}{\rho_{0}}\right)$, $K$ is the 
classical bulk modulus, $S_{1}$, $S_{2}$, $S_{3}$ are the 
intercept of the $U$-$u_{p}$ curve. Parameters $c$, $S_{1}$, 
$S_{2}$, $S_{3}$, $\gamma_{0}$, $a$ represent material properties 
which define its EOS. Parameters have been defined to cover a 
large number of isotropic materials \cite{Steinberg}.

\section{An anisotropic Equation of State}
\label{sec:2}

Before discussing an anisotropic equation of state, the
generalized decomposition of the stress tensor is summarized
in \cite{Lukyanov}. The generalized decomposition framework will
provide a useful point of construction an anisotropic equation of
state.

\subsection{Generalized decomposition of the stress tensor:
$\alpha$-$\beta$ decomposition}
\label{sec:3}

The definition of pressure in the case of an anisotropic solids
should be the result of stating that the "pressure" term should
only produce a change of scale, i.e. isotropic state of strain.
This should allow the calculation of the direct stress ratios that
will produce only a change of scale \cite{Lukyanov}. The
generalized decomposition of the stress tensor $\sigma_{ij}$ is defined as:
\begin{equation}
\label{Luk3_1} \displaystyle \tilde{{\rm P}}:\tilde{S}= 0 \quad \mbox{or}\quad
\alpha_{ij}\tilde{S}_{ij}=0, \quad \sigma_{ij}=- p^{\ast}
\alpha_{ij} + \tilde{S}_{ij},
\end{equation}
where $\tilde{{\rm P}}=p^{\ast}\alpha_{ij}$ is the generalized spherical part
of the stress tensor, $\tilde{S}=\tilde{S}_{ij}$ is the generalized deviatoric
stress tensor, $p^{\ast}$ is the total generalized "pressure" and $\alpha_{ij}$
is the first generalization of the Kronecker's delta symbol. The constructive
definition of the tensor $\alpha_{ij}$ is based on the fact that stress tensor
$p\alpha_{ij}$ is induced in the anisotropic medium by the applied
isotropic strain tensor $\displaystyle\frac{\varepsilon_{v}}{3}\delta_{ij}$, i.e.
\begin{equation}
\label{Luk3_1_1} \displaystyle
p\alpha_{ij}=-\frac{\varepsilon_{v}}{3}C_{ijkl}\delta_{kl}, \quad
p=-K_{C}\varepsilon_{v}
\end{equation}
where $p$ is the pressure, $\varepsilon_{v}$ is the volumetric strain,
$\delta_{ij}$ is the Kronecker's delta symbol (unit tensor),
$C_{ijkl}$ is the elastic stiffness matrix, and $K_{C}$ is the first
generalized bulk modulus. The expressions for $\alpha_{ij}$ components
and $K_{C}$ are presented at the end of Sect.~\ref{sec:3}. Besides,
everywhere contraction by repeating indexes is assumed. Using (\ref{Luk3_1}),
the following expression for total generalized "pressure"
$p^{\ast}$ can be obtained:
\begin{equation}
\label{Luk3_2} \displaystyle p^{\ast}=
-\frac{{\sigma_{ij}\alpha_{ij}}}{{\alpha_{kl}\alpha_{kl}}}=
-\frac{{1}}{{\left\|{\alpha}\right\|^{2}}}\sigma_{ij}\alpha _{ij},
\end{equation}
where $\left\|{\alpha}\right\|^{2}=\alpha_{ij}\alpha_{ij}
=\alpha_{11}^{2}+\alpha_{22}^{2}+\alpha_{33}^{2}$, and finally,
the expression for the generalized deviatoric part
of the stress tensor can be rewritten in the following form:
\begin{equation}
\label{Luk3_3}\displaystyle
\tilde{S}_{ij}=\sigma_{ij}-\alpha_{ij}\cdot
\frac{{1}}{{\left\|{\alpha}\right\|^{2}}}\sigma_{kl}\alpha_{kl} .
\end{equation}
For anisotropic materials, the total generalized "pressure"
$p^{*}$ has been expressed \cite{Lukyanov} as:
\begin{equation}
\label{Luk3_4}\displaystyle p^{*}=p+\frac{{\beta_{ij}\tilde
{S}_{ij}}}{{\beta_{kl}\alpha_{kl}} }, \
p = -\frac{\beta_{ij}\sigma_{ij}}{\beta_{ij}\alpha_{ij}},
\end{equation}
where $p$ is the pressure related to the volumetric deformation
(\ref{Luk3_1_1}) and $\beta_{ij}$ is the second generalization of the
Kronecker's delta symbol. The constructive definition of the tensor
$\beta_{ij}$ is based on the fact that stress tensor $p\beta_{ij}$ is
independent of the stress tensor $C_{ijkl}e_{kl}$, i.e. their contraction
product is zero for any deviatoric strain tensor $e_{kl}$, where $C_{ijkl}$
is the elastic stiffness matrix. The following relation describes the
functional definition of the second order material tensor $\beta_{ij}$:
\begin{equation}
\label{Luk3_4_1}\displaystyle \beta_{ij}C_{ijkl}=3K_{S}\delta_{kl},
\end{equation}
where $K_{S}$ represents the second generalized bulk modulus. The solution of
equations (\ref{Luk3_4_1}) in terms of $\beta_{ij}$ components and expression
for $K_{S}$ are presented below. Equations (\ref{Luk3_4}) define the correct
generalized "pressure" for the elastic regime. Note that the generalized
decomposition of the stress tensor can be applied for all
anisotropic solids of any symmetry and represents a mathematically
consistent generalization of the conventional isotropic case. The
procedure of construction for the tensor $\alpha_{kl}$ has been
defined in \cite{Lukyanov}. The elements of the tensor $\alpha_{kl}$
are
\begin{equation}
\label{Luk3_5}
\begin{array}{*{20}c}
\displaystyle\alpha _{11}=\left(\sum^{3}_{k=1}C_{k1}\right)\cdot 3\bar {K}_{C}, \
\alpha _{22}=\left(\sum^{3}_{k=1}C_{k2}\right)\cdot 3\bar {K}_{C}, \hfill \\
\displaystyle\alpha _{33}=\left(\sum^{3}_{k=1}C_{k3}\right)\cdot 3\bar {K}_{C}, \
\alpha_{ij}\alpha_{ij}=3,\hfill
\end{array}
\end{equation}
\begin{equation}
\label{Luk3_6}
\begin{array}{*{20}c}
\displaystyle K_{C} =
\frac{1}{3\sqrt{3}}
\sqrt{\left(\sum^{3}_{k=1}C_{k1}\right)^{2}+
\left(\sum^{3}_{k=1}C_{k2}\right)^{2}+
\left(\sum^{3}_{k=1}C_{k3}\right)^{2}},\hfill \\
\displaystyle K_{C} = \frac{1}{9\bar{K}_{C}}, \hfill
\end{array}
\end{equation}
where $C_{ij}$ is the elastic stiffness matrix (written in
Voigt notation). The elements of the tensor
$\beta_{kl}$ are
\begin{equation}
\label{Luk3_7}
\begin{array}{*{20}c}
\displaystyle \beta_{11} =\left(\sum^{3}_{k=1} J_{k1}\right)\cdot 3K_{S}, \
\beta_{22} =\left(\sum^{3}_{k=1} J_{k2}\right)\cdot 3K_{S}, \hfill \\
\displaystyle \beta_{33} =\left(\sum^{3}_{k=1} J_{k3}\right)\cdot 3K_{S}, \
\beta_{ij}\beta_{ij}=3, \hfill
\end{array}
\end{equation}
\begin{equation}
\label{Luk3_8}
\displaystyle\frac{1}{K_{S}}=
\sqrt{3}\sqrt{\left(\sum^{3}_{k=1}J_{k1}\right)^{2}
+\left(\sum^{3}_{k=1}J_{k2}\right)^{2}+
\left(\sum^{3}_{k=1}J_{k3}\right)^{2}},
\end{equation}
where $J_{ij}$ are elements of compliance matrix (written in Voigt
notation). In the limit of isotropy, the proposed generalization
returns to the traditional classical case where tensors
$\alpha_{ij}$, $\beta_{ij}$ equal $\delta_{ij}$ and equations
(\ref{Luk3_4}) take the form:
\begin{equation}
\label{Luk3_9} p^{\ast}=-\frac{{\sigma_{ij}\delta_{ij}}}{{\delta
_{kl}\delta_{kl}}}=-\frac{{1}}{{3}}\sigma_{kk},\quad p = -
\frac{{\beta_{ij}\sigma_{ij}}}{{\beta_{kl}\alpha_{kl}}}= -
\frac{{1}}{{3}}\sigma_{kk},
\end{equation}
where $\displaystyle p=-\frac{\sigma_{kk}}{3}$ is the classical
hydrostatic pressure. Also, parameters $K_{C}$ and $K_{S}$ were
considered as the first and the second generalized bulk moduli. In
the limit of isotropy they reduce to the well-know expression for
conventional bulk modulus.

\subsection{Thermodynamical framework}
\label{sec:4}

During shock loading, the medium undergoes nonlinear behavior;
therefore an equation of state (EOS) is required to describe the
medium's response under shock loading conditions. It is convenient
for use in numerical codes to have an analytical form of the EOS.
Such an analytic form is at best an approximation to the true
relationship. Thermodynamical definition of the Gr\"{u}neisen
parameter $\Gamma (\nu, T)$ is the following:
\begin{equation}
\label{Luk4_1} \Gamma(\nu,
T)=\frac{\nu}{C_{\nu}}\left(\frac{\partial S}{\partial
\nu}\right)_{T},
\end{equation}
where $S$ is the entropy, $T$ is the temperature, $C_{\nu}$ is the
heat capacity at constant specific volume. This definition may be
generalized to a tensor $\Gamma_{ij}$ for the stress-strain
variables \cite{Wallace}:
\begin{equation}
\label{Luk4_2} \Gamma_{ij}=
\frac{\nu}{C_{\varepsilon}}\left(\frac{\partial S}{\partial
\varepsilon_{ij}}\right)_{T},
\end{equation}
where $\varepsilon_{ij}$ is the strain tensor and
$C_{\varepsilon}$ is the heat capacity at constant strains. Using
Maxwell's thermodynamic relations, the isotropic Mie-Gr\"{u}neisen
EOS can be generalized for anisotropic media as follows:
\begin{equation}
\label{Luk4_3} \Gamma_{ij}=
\nu\left(\frac{\partial \tau_{ij}}{\partial e}\right)_{\varepsilon_{ij}},
\end{equation}
where $\tau_{ij}$ is an anisotropic stress tensor produced by the
isotropic strain state. Also, this EOS (\ref{Luk4_3}) is
consistent with the Wallace's model \cite{Wallace}, where a full
anisotropic stress was used. It has been shown (see
Sect.~\ref{sec:3}) that the stress tensor $p\alpha_{ij}$ is
produced by the spherical isotropic strain tensor
$\varepsilon\delta_{ij}$, hence (\ref{Luk4_3}) can be re-written
as:
\begin{equation}
\label{Luk4_4} \Gamma_{ij}=\alpha_{ij}\Gamma, \
\Gamma=\nu\left(\frac{\partial p}{\partial e}\right)_{\nu},
\end{equation}
where the pressure $p$ is defined by the constitutive equations
(\ref{Luk3_4}) and $\varepsilon=\frac{1}{3}\varepsilon_{v}$
is the isotropic strain.

\subsection{An anisotropic EOS}
\label{sec:5}

The equations (\ref{Luk3_4}) define the correct generalized
"pressure" for the elastic regime. To provide an appropriate
description for general anisotropic materials behavior at high
pressures, the pressure $p$ related to the volumetric deformations
is described by $p^{EOS}$ (\ref{Luk1_4}), (\ref{Luk4_4}).
Therefore, an appropriate description of generalized hydrostatic
"pressure" at high pressures has the following form:
\begin{equation}
\label{Luk5_2}\displaystyle p^{*}=p^{EOS}+\frac{{\beta_{ij}\tilde
{S}_{ij}}}{{\beta_{kl}\alpha_{kl}} }, \
\sigma_{ij} = -p^{*}\alpha_{ij}+ \tilde{S}_{ij},
\end{equation}
which also describes correctly the medium's behavior at small
volumetric strains. To be consistent with the definition of the
isotropic bulk speed of sound, the following definitions of the
first $c_{I}$ and the second $c_{II}$ bulk speed of sound for
anisotropic medium are assumed:
\begin{equation}
\label{Luk5_3} \displaystyle c_{I}=\sqrt\frac{K_{C}}{\rho_{0}},
\ c_{II}=\sqrt\frac{K_{S}}{\rho_{0}},
\end{equation}
where the generalized bulk moduli $K_{C}$, $K_{S}$ are defined
according to (\ref{Luk3_6}) and (\ref{Luk3_8}) respectively.
Parameters $c\in [c_{II},c_{I}]$, $S_{1}$, $S_{2}$, $S_{3}$,
$\gamma_{0}$, $a$ represent material properties which define its
EOS (\ref{Luk1_4}). A description of their numerical values for
AA7010-T6 is shown in Table \ref{tab:1}.
\begin{table}
\caption{EOS data for AA7010-T6 used in analysis.}
\label{tab:1}       
\begin{tabular}{lll}
\hline\noalign{\smallskip}
Parameter & Description & AA7010-T6  \\
\noalign{\smallskip}\hline\noalign{\smallskip}
$c \ \left[m/s\right]$ & Velocity curve intercept & 5154 \\
$S_{1}$ & First slope coefficient & 1.4 \\
$S_{2}$ & Second slope coefficient & 0.0 \\
$S_{3}$ & Third slope coefficient &  0.0 \\
$\gamma_{0}$ & Gr\"{u}neisen gamma & 2.0  \\
$a$ & First-order volume correction & 0.48  \\
$\rho_{0} \ \left[kg/m^{3}\right]$ & Initial density & 2810  \\
$K_{C} \ \left[GPa\right]$ &  Generalized bulk modulus & 74.65  \\
\noalign{\smallskip}\hline
\end{tabular}
\end{table}
\section{A non-associated anisotropic plasticity model}
\label{sec:6}

The main aspects of a phenomenological strength model can be
characterized by a yield criterion representing a surface that
separates the elastic and plastic regions of the stress space, a
flow potential gradient that represents the direction of plastic
strain rate, a strain hardening rule and that plastic flow is
incompressible.

\subsection{An anisotropic yield surface}
\label{sec:7}

Following the research of Spitzig and Richmond
\cite{SpitzigRichmond}, and Stoughton and Yoon
\cite{StoughtonYoon}, the mathematically consistent yield function
of a fully anisotropic material based on generalized decomposition
of the stress tensor is developed :
\begin{equation}
\label{Luk7_1}
\begin{array}{*{20}c}
\displaystyle \hat{F}\left({\tilde {S}_{ij}}\right) =
\Psi\left({\tilde {S}_{ij}}\right)\left( {1 + \chi p^{\ast} }
\right) \le Y\left( {\bar {\varepsilon
}_{p}}\right), \hfill \\
\displaystyle p^{\ast} =
\frac{{1}}{{\left\| {\alpha}  \right\|^{2}}}\sigma _{ij}
\alpha _{ij}, \hfill
\end{array}
\end{equation}
where $\Psi\left({\tilde {S}_{ij}}\right)$ is described
by generalized Hill's yield functions:
\begin{equation}
\label{Luk6_2}
\displaystyle
\begin{array}{l}
\Psi^{2}\left({\tilde {S}_{ij}}  \right) = F\left( {\alpha _{3}
\tilde {S}_{2} - \alpha _{2} \tilde {S}_{3}}  \right)^{2} + G\left(
{\alpha _{1} \tilde {S}_{3} - \alpha _{3} \tilde {S}_{1}}  \right)^{2}
+ \hfill \\
+H\left( {\alpha _{22} \tilde {S}_{11} - \alpha _{11} \tilde {S}_{22}}
\right)^{2}+ 2N\tilde {S}_{12}^{2} + 2L\tilde {S}_{23}^{2}
+ 2M\tilde {S}_{13}^{2} \hfill
\end{array},
\end{equation}
where $\tilde {S}_{ij}$ is the generalized deviatoric stress
tensor ($\tilde{S}_{i}=\tilde{S}_{ii}, i=1,2,3$); $\alpha_{ij}$ is
the generalized Kronecker's symbol \cite{Lukyanov}
($\alpha_{i}=\alpha_{ii}, i=1,2,3$). The material constants
$\chi$, $F$, $G$, $H$, $N$, $L$, $M$ are specified in terms of
selected initial yield stresses in uniaxial tension, compression,
and equibiaxial tension. It is important to note that plasticity
model (\ref{Luk6_2}) is naturally independent from the generalized
hydrostatic pressure and therefore, the following equality can be
written:
\begin{equation}
\label{Luk7_3}
\Psi\left({\tilde {S}_{ij}}\right)\equiv\Psi\left({\sigma _{ij}}\right),\quad
\sigma_{ij} = p^{\ast}\alpha _{ij} + \tilde {S}_{ij}, \ p^{\ast}  \ne 0,
\end{equation}
where $\sigma _{ij}$ is the stress tensor. In this paper, a
uniaxial strain state (one-dimensional reduced mathematical
formulation in strain space) and the adiabatic approximation
assumptions are considered for modelling shock waves propagation
in anisotropic solids. Therefore, material parameters and their
numerical values for AA7010-T6 are taken as presented in Table
\ref{tab:2}.
\begin{table}
\caption{Material properties for AA7010-T6 used in analysis.}
\label{tab:2}       
\begin{tabular}{lll}
\hline\noalign{\smallskip}
Parameter & Description & AA7010-T6  \\
\noalign{\smallskip}\hline\noalign{\smallskip}
$F$ & Anisotropy coefficient& 0.6898 \\
$G$ & Anisotropy coefficient& 0.2873 \\
$H$ & Anisotropy coefficient& 0.6824 \\
$Y \ [MPa]$ & Yield stress& 500.0 \\
$\chi \ \left[MPa^{-1}\right]$ & Pressure dependency factor & 0.0  \\
$\alpha_{11}$ & Tensor $\alpha_{ij}$ (11 direction) & 0.9976  \\
$\alpha_{22}$ & Tensor $\alpha_{ij}$ (22 direction) & 1.0029  \\
$\alpha_{33}$ & Tensor $\alpha_{ij}$ (33 direction) & 0.9994  \\
\noalign{\smallskip}\hline
\end{tabular}
\end{table}
\subsection{An anisotropic plastic potential}
\label{sec:8}

A flow potential gradient that represents the direction of plastic
strain rate is described by the classical Hill's anisotropic
plastic potential:
\begin{equation}
\label{Luk8_1}
\displaystyle
\begin{array}{l}
\displaystyle\Pi\left(\sigma_{ij}, \gamma_{ij}\right)=
\left[\bar{F}\left({\sigma_{22}-\sigma_{33}}\right)^{2}+\bar{G}\left(
{\sigma_{33}-\sigma_{11}}\right)^{2}+\right. \hfill \\
\displaystyle\left. + \bar{H}\left( {\sigma_{11} - \sigma_{22}}\right)^{2}
+ 2\bar{N}\sigma_{12}^{2} + 2\bar{L}\sigma_{23}^{2} + 2\bar{M}\sigma_{13}^{2}\right]^{1/2}
-\sigma_{\Pi} \hfill \\
\displaystyle D^{p}_{ij}=\dot{\lambda}\frac{\partial \Pi\left(\sigma_{kl},
\gamma_{kl}\right)}{\partial \sigma_{kl}}
\hfill
\end{array},
\end{equation}
and has a strain-rate dependent constant $\sigma_{\Pi}$ including
hardening modulus for plastic potential; $\gamma_{ij}$ is the back
stress for plastic potential, $D^{p}_{ij}$ is the plastic strain
rate tensor. It is assumed that the parameters $\bar{F}$, $\bar{G}$,
$\bar{H}$, $\bar{N}$, $\bar{L}$, $\bar{M}$ can be determined from 
the anisotropy parameters $R$, $P$, $Q_{23}$, $Q_{31}$ and $Q_{12}$
as for the Hill's yield function \cite{Hill}. The constitutive equations
are integrated using the tangent stiffness matrix. The numerical values
for plastic potential parameters were taken from \cite{DeVuyst}.

\section{Simulation of anisotropic shock wave propagation}
\label{sec:9}

The plane shock-wave technique provides a powerful tool for
studying material properties at different strain rates
\cite{Steinberg}, \cite{Meyers},
\cite{BourneGray},\cite{Millettetal1}. The characteristics as
spall pressure, shock velocity, particle velocity, Hugoniot
elastic limit, thickness of the spall section, time to spall, and
free surface velocity of the spall section can be measured and
used for the characterization of material dynamic response
\cite{BourneGray}, \cite{Millettetal1}, \cite{Stoffel},
\cite{MeziereMillettBourne}.

\subsection{Description of Experiment}
\label{sec:10}

The flyer plates were launched using a 75 mm bore, 1 m long and 50
mm bore, and 1 m long single stage gas gun the Royal Military
College of Science \cite{BourneStevens}. The shock propagation in
the target is monitored using manganin stress gauges, placed
between target plate and PMMA ( poly methylmethacrylate) plate
within the target assembly. 
The results from the stress gauges were converted by N. Bourne and
J. Millett to in material (Target) values $\sigma_{M}$, using the
shock impedances of the target $A_{T}$ and PMMA $A_{P}$, via the
well-known relation
\begin{equation}
\label{Luk10_1}
\displaystyle\sigma_{M}=\frac{A_{T}+A_{P}}{2A_{P}}\sigma_{P},
\end{equation}
where $\sigma_{P}$ is the stress gauges values. The 2.5mm thick
flyer plates of 6082-T6 (dural) were impacted onto the targets
(test material AA 7010-T6) over the velocity range 234 to 895,
inducing stresses in the range 2.7 to 7.2 GPa. The aluminium alloy
6082-T6 was chosen as the flyer due to the close similarity in
density and wave speeds, so that the impact experiments were near
symmetrical \cite{DeVuyst}. The elastic material properties of
7010-T6 can be found in \cite{DeVuyst}. Material properties of
plates 6082-T6 and PMMA can be found in \cite{Steinberg}.

Using a finite-difference wave propagation code
\cite{KiselevLukyanov}, numerical simulations of plate-impact test
were performed with a 5 mm thick target plate, 2.5 mm thick flyer
plate and 5 mm PMMA plate. Based on the characteristics of this
plate impact problem, the plates (numerical domains), which are
used in the numerical simulation, are modelled as 1D bars
\cite{KiselevLukyanov}. The mesh resolutions were sufficient to
allow the resolution of all the relevant elastic and plastic waves
in the target and flyer. The stress time histories were recorded
in the middle of the target plate and at the back of the test
specimen (the first FD element in the PMMA connected to test
plate).

\subsection{Shock wave propagation in 7010 T6 anisotropic
aluminum alloy}\label{sec:11}

The experimental data for AA7010 T6 presented here correspond to
the plate impact test performed by N. Bourne and J. Millett at
Royal Military College of Science (published in \cite{DeVuyst})
with impact velocities of 450 m/s and 895 m/s.
%
%
\begin{figure}
\includegraphics[width=3.0in,height=2.5in]{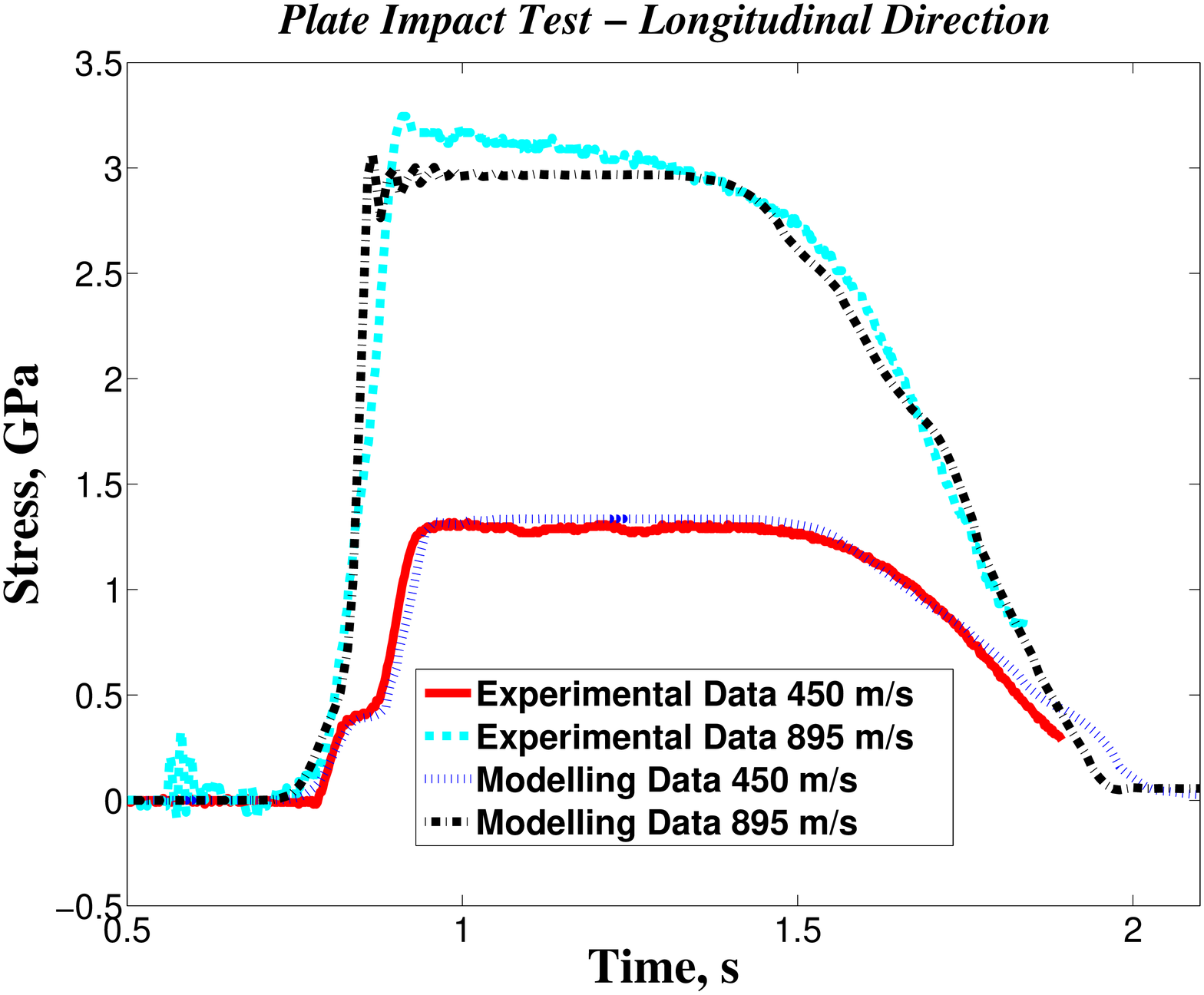}
\caption{ Back-surface gauge stress traces from  plate-impact experiments
versus numerical simulation of  stress (PMMA) waves for plate impact test
(impact velocity 450 m/s and 895 m/s) - aluminum alloy AA7010-T6 (Longitudinal
Direction).}
\label{fig:2}       
\end{figure}
%
%
\begin{figure}
\includegraphics[width=3.0in,height=2.5in]{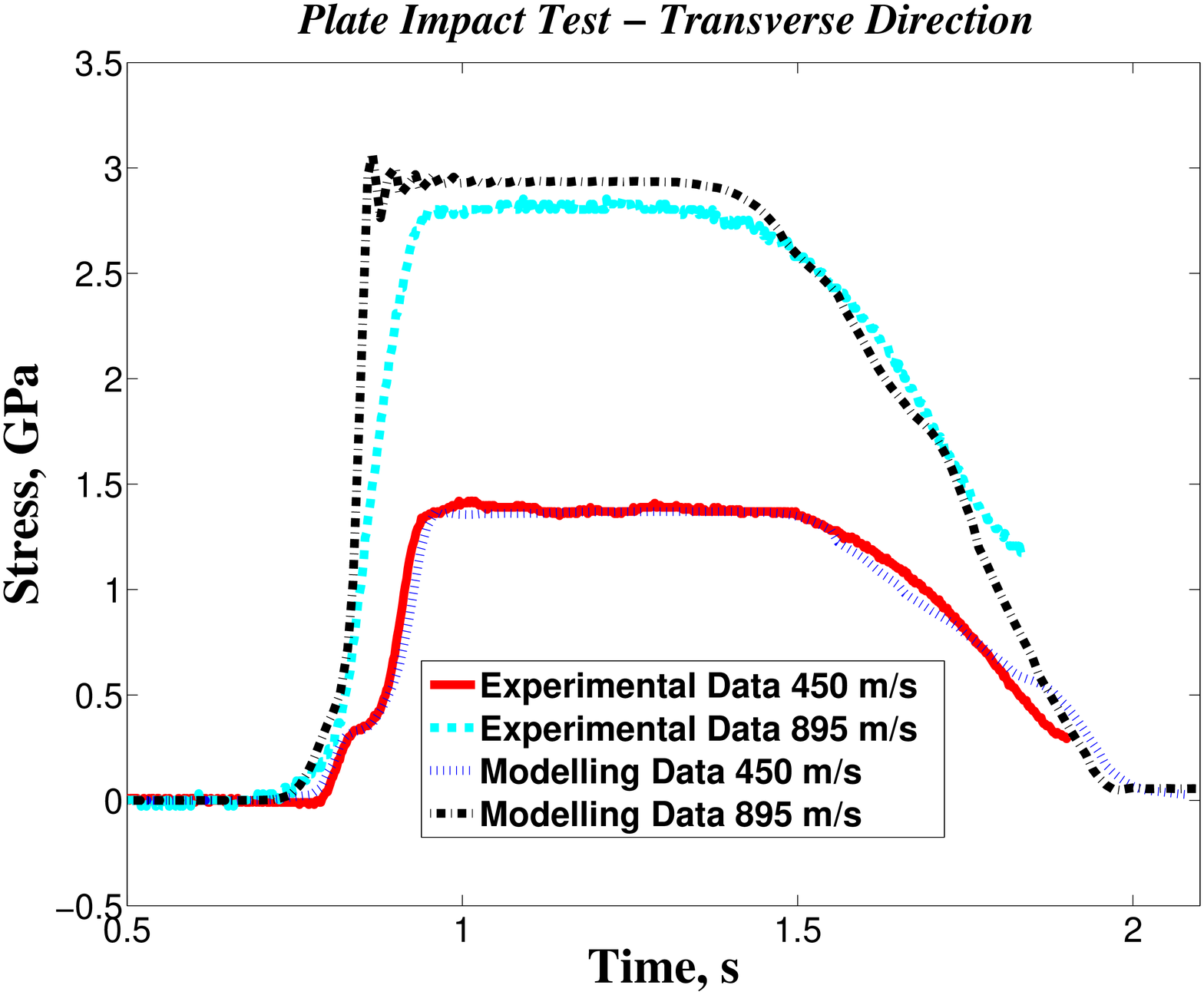}
\caption{ Back-surface gauge stress traces from  plate-impact experiments
versus numerical simulation of  stress (PMMA) waves for plate impact test
(impact velocity 450 m/s and 895 m/s) - aluminum alloy AA7010-T6 (Transverse
Direction).}
\label{fig:3}       
\end{figure}
Figures \ref{fig:2} and \ref{fig:3} show the comparison between experimental data and the
numerical simulation resulting from the new anisotropic equation of state
and non-associated anisotropic plasticity model for the longitudinal
and transverse cases.

The experimental values, 0.39 GPa and 0.33 GPa for elastic
response from the longitudinal and short transverse directions,
respectively figures \ref{fig:2} and \ref{fig:3}, are in good correlation with the
modelled values of the HEL longitudinal - 0.395 GPa and short
transverse - 0.333 GPa. The errors with respect to the
experimental values are 1.4 \% and 0.9 \%, respectively to the
longitudinal and short transverse directions. Further comparison
shows that the stress pulse width (approximately, 0.81 $\mu s$ )
and release trace are in good agreement with the experiment as
well (see fig. \ref{fig:2} and \ref{fig:3}). Besides, another important
characteristic, the arrival time to the HEL and the plastic wave
velocity are in good correlation with experimental data.  The main
conclusion obtained from these results is that the non-associated
anisotropic plasticity model, as it stands, is suitable for
simulating elastoplastic shock wave propagation in anisotropic
solids. Different HELs are obtained when the material is impacted
in different directions; their excellent agreement with the
experiment demonstrates adequateness of the proposed anisotropic
plasticity model. However, further work is required both in the
experimental and constitutive modelling areas to find a better
description of anisotropic material behavior and, particulary,
Hugoniot stress level.

\section{Conclusions}\label{sec:12}

The EOS proposed in this paper represents a physical
generalization of the classical Mie-Gr\"{u}neisen equation of
state for isotropic materials. Based on the $\alpha-\beta$
generalized decomposition of stress tensor, the modified
Mie-Gr\"{u}neisen equation of state combined with non-associated
plasticity model forms a system of constitutive equations suitable
for shock wave propagation in single crystals and polycrystalline
alloys. The behavior of the aluminum alloy 7010-T6 under shock
loading conditions was investigated. A comparison of the
experimental HELs with numerical simulation shows an excellent
agreement (maximum error is 1.4\%). Furthermore, the good
agreement of the general pulse shape and Hugoniot stress level
suggests that the EOS is performing satisfactorily. The
significance of the proposed strength model includes also the
distortion of the yield function shape in tension, compression and
in different principal directions of anisotropy (e.g., $0^{0}$,
$90^{0}$) which can be used to describe the anisotropic strength
differential effect (anisotropic SDE). However, further
development of the constitutive equations taking into account
strain rate sensitivity is required.

\section{Acknowledgments}
\label{sec:13}

Author thanks Prof. V. Penjkov for many useful suggestions
regarding this work.  The discussions regarding the shock waves
experiments with Prof. N. K. Bourne and Dr. J. C. F. Millett
during the meetings at Cranfield University are also greatly
appreciated.
%

\end{document}